\def\year{2017}\relax

\documentclass[letterpaper]{article}
\usepackage{aaai17}
\usepackage{times}
\usepackage{graphicx}
\usepackage{subcaption}
\usepackage{helvet}
\usepackage{courier}
\usepackage{color}
\usepackage{amsfonts}
\usepackage{amsmath}
\usepackage{url}
\usepackage{booktabs}

\definecolor{wine}{rgb}{0.52,0.011,0.13}

\newcommand{\spec}{{\mathcal{S}}}
\newcommand{\vol}{{\mathcal{V}}}
\newcommand{\niche}{{\mathcal{N}}}
\newcommand{\dynam}{{\mathcal{D}}}
\newcommand{\prob}{{\text{P}}}

\newcommand{\cidentity}{{community identity}}

\newcommand{\xhdr}[1]{{\noindent\bfseries #1.}}

\newcommand{\cut}[1]{}

\newcommand*\samethanks[1][\value{footnote}]{\footnotemark[#1]}

\title{Community Identity and User Engagement in a Multi-Community Landscape}
\author{
Justine Zhang{\Large \thanks{The two first authors contributed equally and are ordered non-alphabetically to balance ordering in another collaboration.}}\\
      {Cornell University}\\
      \texttt{jz727@cornell.edu}
\And
William L. Hamilton{\Large \samethanks[1]}\\
Stanford University\\
\texttt{wleif@stanford.edu}
\AND
\hspace{0.1 in} Cristian Danescu-Niculescu-Mizil  \\
Cornell University \\
\texttt{cristian@cs.cornell.edu}
\And
Dan Jurafsky\\
{Stanford University}\\
\texttt{jurafsky@stanford.edu}
\And
Jure Leskovec\\
{Stanford University}\\
\texttt{jure@cs.stanford.edu}
}

\begin{document}
\maketitle
\begin{abstract}

A community's identity defines and shapes its internal dynamics. 
Our current understanding of this 
interplay
 is mostly limited to glimpses gathered from isolated studies of individual communities. 
In this work we provide a systematic exploration of the nature of this relation across a wide variety of online communities. 
To this end we introduce a quantitative, language-based typology reflecting two key aspects of a community's identity: how \textit{distinctive}, and how temporally %
\textit{dynamic} it is.
By mapping almost 300 Reddit communities
into the landscape induced by this typology, we reveal regularities in how patterns of user engagement vary with the characteristics of a community.

Our results suggest that 
the way
 new and existing users engage with a community depends strongly and systematically on the nature of the collective identity it fosters, in ways that are highly consequential to community maintainers.
For example, communities with distinctive and highly dynamic identities are more likely to retain their users.  However, such niche communities also exhibit much larger acculturation gaps between existing users and newcomers, which potentially hinder the integration of the latter. %

 More generally, our methodology reveals differences in how various social phenomena manifest across communities, and shows that structuring the multi-community landscape can lead to a better understanding of the systematic nature of this diversity.

\end{abstract}

\section{Introduction}

\begin{quote}
\footnotesize
\em
``If each city is like a game of chess, the day when I have learned the rules, I shall finally possess my empire, even if I shall never succeed in knowing all the cities it contains.''
\end{quote}
\vspace{-0.2cm}
\begin{flushright}
\footnotesize
--- Italo Calvino,  Invisible Cities
\end{flushright}

A community's identity---defined through the common interests and shared experiences of its users---shapes various facets of the social dynamics within it \cite{ren_applying_2007,tajfel_social_2010,ren_encouraging_2012}.
Numerous instances of this interplay between a community's identity and social dynamics have been extensively studied in the context of individual online communities \cite{bryant2005becoming,lampe2010motivations,danescu-niculescu-mizil_no_2013}.
However, the sheer variety of online platforms complicates the task of generalizing insights 
 beyond these isolated, single-community glimpses.  A new way to reason about the variation across multiple communities is needed in order to systematically characterize the relationship between %
properties of a community
  and the dynamics taking place within.

One especially important component of community dynamics is user 
engagement.
We can aim to %
understand why users join certain communities \cite{panciera_wikipedians_2009}, what factors influence user retention \cite{dror_churn_2012}, and how users react to innovation \cite{danescu-niculescu-mizil_no_2013}. While striking patterns of user engagement have been uncovered in prior case studies of 
individual
 communities \cite{postmes_formation_2000,huffaker2006computational,fugelstad2012makes,Otterbacher:ProceedingsOfCscw:2012,mcauley_amateurs_2013}, 
we do not know whether 
these observations hold beyond these cases,
or when we can draw analogies between different communities.  Are there certain types of communities where we can expect similar 
or contrasting engagement patterns?
To address such questions quantitatively
we need to provide 
structure to 
the diverse and complex 
space of online communities. 
Organizing the multi-community landscape would allow us to both 
characterize
 individual points within this space, and reason about 
 systematic 
 variations in patterns
 of user engagement 
  across the space.

\xhdr{Present work: Structuring the multi-community space}
In order to systematically understand the relationship between community identity\footnote{We use ``community identity'' and ``collective identity" 
 interchangeably to refer to the shared definition of a group, derived from members' common interests and shared experiences.
We are not directly concerned with the more sociopolitical and psychological connotations of these terms \cite{polletta2001collective,simon2001politicized,ashmore2004organizing}.}%
and user engagement we introduce a quantitative typology of online communities.
Our typology is based on two key aspects of \cidentity: 
how \textit{distinctive}---or niche---a community's interests are relative to other communities, and how \textit{dynamic}---or volatile---these interests are over time.
These axes aim to capture 
the salience of a community's identity and dynamics of its temporal evolution. 

Our main insight in implementing this typology automatically and at scale is that the {\em language} used within a community can 
simultaneously capture how 
distinctive and dynamic its interests are. 
This 
language-based
 approach draws on a wealth of literature characterizing linguistic variation in online communities and its relationship to community and user identity \cite{cassell_language_2005,danescu-niculescu-mizil_no_2013,bamman2014gender,tran2016characterizing,eisenstein2017geographical}.
 Basing our typology on language is also convenient since it renders our framework immediately applicable to a wide variety of online communities, where 
communication is primarily recorded in a textual format. 
Using our framework, we map almost 300 Reddit communities onto the landscape defined by the two axes of our typology (Section \ref{sec:typology}). 
We find that this 
mapping
induces conceptually sound categorizations that effectively capture key aspects of community-level social dynamics. 
In particular, we quantitatively validate the effectiveness of our 
mapping
 by showing that 
our two-dimensional typology 
encodes signals that are predictive of community-level rates of user retention, complementing strong activity-based features.

\xhdr{Engagement and community identity} We apply our framework to understand how two important aspects of user engagement in a community---the community's propensity to retain its users  (Section \ref{sec:engagement}), and its permeability to new members (Section \ref{sec:acculturation})---vary according to the type of collective identity it fosters.
We find that communities that are characterized by specialized, constantly-updating
 content have higher user retention rates, 
but also exhibit
 larger linguistic gaps 
 that separate newcomers from established members.

More closely examining factors that could contribute to this linguistic gap, we find that especially within distinctive communities, established users have an increased propensity to engage with the community's specialized content, compared to newcomers (Section \ref{sec:divides}).  Interestingly, while established members of distinctive communities more avidly respond to temporal updates than newcomers, in more generic communities it is the {\em outsiders} who engage more with volatile content, perhaps suggesting that such content may serve as an entry-point to the community (but not necessarily a reason to stay). 
Such insights into the relation between collective identity and user engagement can be informative to community maintainers seeking to better understand growth patterns within their online communities.
More generally, our methodology stands as an example of how sociological questions can be addressed in a multi-community setting. In performing our analyses across a rich variety of communities, we reveal both the diversity of phenomena that can occur,
as well as the systematic nature of this diversity.
\section{A typology of \cidentity}\label{sec:typology}

A community's identity derives from its members' common interests and shared experiences \cite{ashmore2004organizing,ritzer2007blackwell}.
In this work, we structure the multi-community landscape along these 
  two key dimensions of community identity: how {\em distinctive} a community's interests are, and how {\em dynamic} the community is over time.  

We now proceed to outline our quantitative typology, which maps communities along these two dimensions.
We start by providing an intuition through inspecting a few example communities. We then introduce
a generalizable language-based methodology and use it to map a large set of Reddit communities 
onto the landscape defined by our typology of community identity.

\subsection{Overview and intuition}\label{subsec:fig0}

\begin{figure*}
\includegraphics[width=0.95\textwidth]{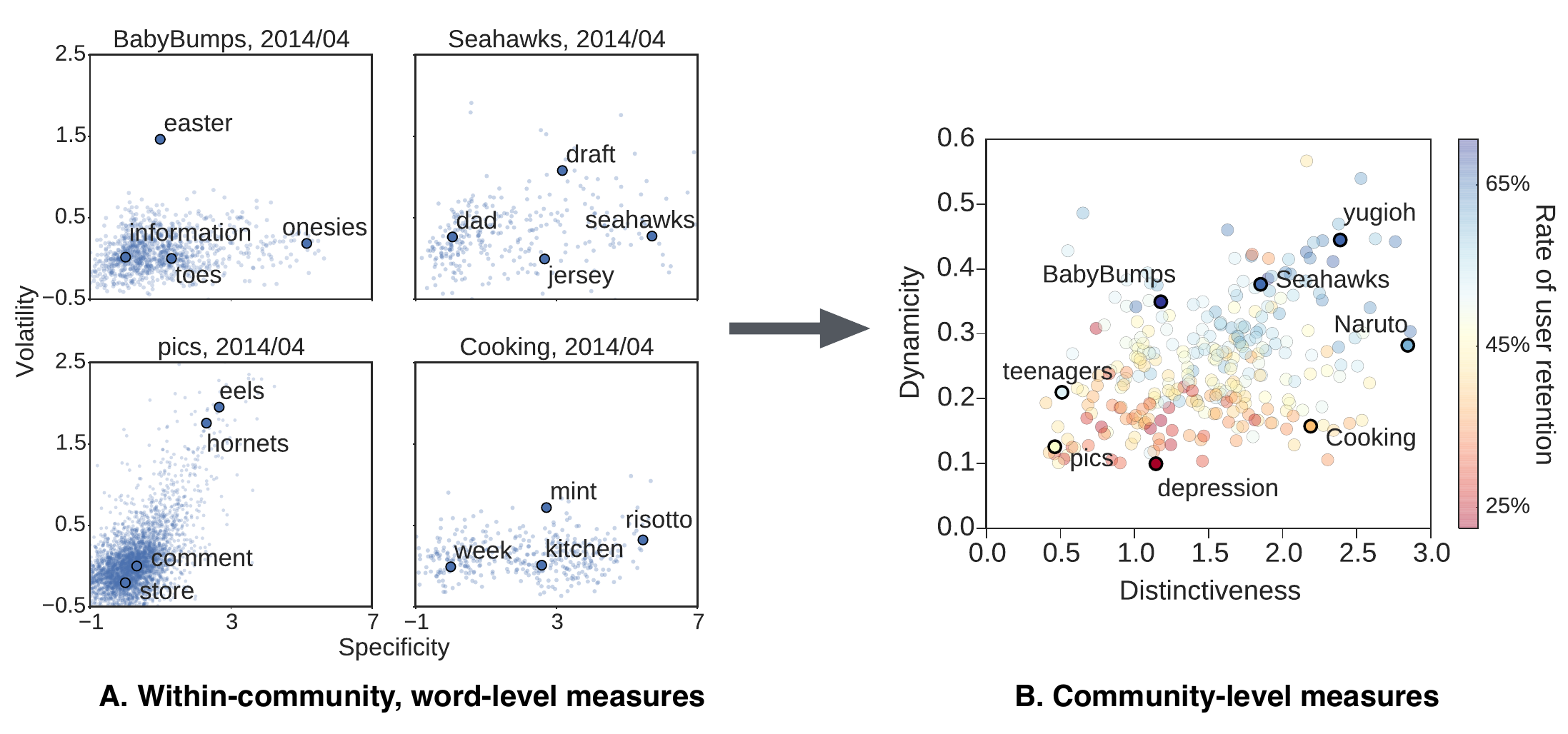}
\vspace{-10pt}
\caption{%
\textbf{A}: Within a community certain words are more community-specific and temporally volatile 
than others. For instance, words like \textit{onesies} are highly specific to the \textit{BabyBumps} community (top left), while words like \textit{easter} are temporally ephemeral. 
\textbf{B}: Extending these word-level measures to communities, we can measure the overall distinctiveness and dynamicity of a community, which are  highly associated with user retention rates (colored heatmap; see Section \ref{sec:engagement}). Communities like \textit{Seahawks} (a football team) and \textit{Cooking} use highly distinctive language. Moreover, \textit{Seahawks} uses very dynamic language, as the discussion continually shifts throughout the football season. 
In contrast, the content of \textit{Cooking} remains stable over time, as does the content of \textit{pics}; though these communities do have ephemeral fads, the %
overall themes discussed generally remain consistent. %
}
\label{fig:figure0}
\end{figure*}

In order to illustrate the diversity within the multi-community space, and 
to provide an
 intuition for the underlying 
  structure captured by the proposed typology, we first examine a few example communities and draw attention to some key social dynamics that occur within them.

We consider four communities from Reddit: in {\em Seahawks}, fans of the Seahawks football team gather to discuss games and players; in {\em BabyBumps}, expecting mothers trade advice and updates on their pregnancy; {\em Cooking} consists of recipe ideas and general discussion about cooking; while in {\em pics}, users share various images of random things (like eels and hornets). We note that these communities are topically contrasting and foster fairly disjoint user bases. 
Additionally, these communities exhibit varied patterns of user engagement. While Seahawks maintains a devoted set of users from month to month, pics is dominated by transient users who post a few times and then depart. 

Discussions {\em within} these communities also span varied sets of interests. Some of these interests
are more specific to the community than others: {\em risotto}, for example, is seldom a discussion point beyond Cooking.
Additionally, some interests consistently recur, while others are specific to a particular time: {\em kitchen}s are a consistent focus point for cooking, but {\em mint} is only in season during spring. 
Coupling
specificity and consistency
 we find interests such as {\em easter}, which isn't particularly specific to BabyBumps but gains prominence in that community around Easter (see Figure \ref{fig:figure0}.A for further examples).

These specific interests provide a window into the nature of the communities' interests as a whole, and by extension 
their community identities.
Overall, discussions in Cooking focus on topics which are highly distinctive and consistently recur (like {\em risotto}). 
In contrast, discussions in Seahawks are highly dynamic,
rapidly shifting over time as new games occur and players are traded in and out. 
In the remainder of this section we formally introduce a methodology for mapping communities in this space defined by their {\em distinctiveness} and {\em dynamicity} (examples in Figure~\ref{fig:figure0}.B).%

\subsection{Language-based formalization} 

 Our approach follows the intuition that a distinctive community will use language that is 
 particularly
  {\em specific}, or unique, to that community.
 Similarly, a dynamic community will use {\em volatile} language that rapidly changes across successive windows of time.
To capture this intuition
automatically,
 we 
 start by defining word-level measures of specificity and volatility. 
We then extend these word-level primitives to characterize entire comments, and the community itself. 
Our characterizations of words in a community are motivated by methodology from prior literature that compares the frequency of a word in a particular setting to its frequency in some background distribution, in order to identify instances of linguistic variation \cite{monroe2008fightin,eisenstein2017geographical}. Our particular framework makes this comparison by way of pointwise mutual information (PMI). 

In the following, we use $c$ to denote one community within a 
set
 $\mathcal{C}$ of communities, and $t$ to denote one time period within the entire history $T$ of $\mathcal{C}$. We account for temporal as well as inter-community variation 
  by computing 
 word-level measures
for each time period of each community's history, $c_t$. 
Given a word $w$ used within a particular community $c$ at time $t$, we define two word-level measures:

\xhdr{Specificity}
We quantify the {\em specificity} $\spec_{c} (w)$ of $w$ to $c$ by calculating the PMI of $w$ and $c$, relative to $\mathcal{C}$, 
$$ \spec_{c} (w)=\log\frac{ \prob_{c} (w)}{\prob_{\mathcal{C}} (w)},$$
 where $\prob_{c}(w)$ is $w$'s frequency in $c$. $w$ is {\em specific} to $c$ if it occurs more frequently in $c$ than in the entire set $\mathcal{C}$, hence distinguishing this community from the rest. A word $w$ whose occurrence is decoupled from $c$, and thus has 
$\spec_{c} (w)$ close to 0, is said to be  {\em generic}.

We compute values of 
$\spec_{c_t} (w)$
 for each time period $t$ in $T$; in the above description we drop the time-based subscripts for clarity.

\xhdr{Volatility}
We quantify the {\em volatility} $\vol_{c_t} (w)$ of $w$ to $c_t$ as the PMI of $w$ and $c_t$ relative to $c_T$, the entire history of $c$: $$\vol_{c_t} (w)=\log\frac{ \prob_{c_t} (w)}{\prob_{c_T} (w)}.$$ A word $w$ is {\em volatile} at time $t$ in $c$ if it occurs more frequently at $t$ than in the entire history $T$, behaving as a fad within a small window of time. A word that occurs with similar frequency across time, and hence has $\vol$ close to 0, is said to be {\em stable}.

\xhdr{Extending to utterances}
Using our word-level primitives, we define the specificity of an utterance $d$ in $c$, $\spec_{c} (d)$ as the average specificity of each word in the utterance. 
The volatility of utterances is defined analogously.
\vspace{0.2in}

\subsection{Community-level measures}
Having described these 
 word-level measures, we now proceed to establish the primary axes of our 
 typology:

\xhdr{Distinctiveness}
A community with a very distinctive identity will tend to have distinctive interests, expressed through specialized language.  Formally, we define the distinctiveness of a community $\niche (c_t)$ as the average specificity of all utterances in $c_t$.
We refer to a community with a less distinctive identity as being {\em generic}.

\xhdr{Dynamicity} 
A highly dynamic community constantly shifts interests from one time window to another, and these temporal variations are reflected in its use of volatile language. Formally, we define the dynamicity of a community $\dynam (c_t)$ as the average volatility of all utterances in $c_t$.
We refer to a community whose language is relatively consistent throughout time as being {\em stable}. 

In our subsequent analyses, we focus mostly on examing the {\em average} distinctiveness and dynamicity of a community over time, denoted $\niche (c)$ and $\dynam (c)$.

\subsection{Applying the typology to Reddit}

We now explain how our typology can be %
applied to the particular setting of Reddit, and describe the overall behaviour of our linguistic axes in this context.

\xhdr{Dataset description}
Reddit is a popular website where users form and participate in discussion-based communities called {\em subreddits}.
Within these communities, users post content---such as images, URLs, or questions---which often spark vibrant lengthy discussions in thread-based comment sections. 

The website contains many highly active
subreddits with
 thousands of active subscribers. These communities span an extremely rich variety of topical interests, 
as represented by the examples described earlier.
 They
  also vary along a rich multitude of structural dimensions, such as the number of users, the amount of conversation and social interaction,
  and the 
  social norms determining which types of content become popular.
 The diversity and scope of Reddit's multicommunity ecosystem make it an ideal landscape in which to closely examine the relation between varying community identities and social dynamics.

Our full dataset consists of all subreddits on Reddit from January 2013 to December 2014,\footnote{\url{https://archive.org/details/2015_reddit_comments_corpus}} 
for which there are at least 500 words in the vocabulary used to estimate our measures, in at least 4 months of the subreddit's history. 
We compute our measures over the comments written by users in a community in time windows of {\em months}, for each sufficiently active month, 
and manually remove communities where the bulk of the contributions are in a foreign language. This results in 283 communities ($c$), for a total of 
4,872
 community-months ($c_t$).\footnote{While we chose these cutoffs on the dataset to ensure robust estimates of the linguistic measures, we note that slight relaxations produce qualitatively similar results in the later analyses.}

\begin{table}
\begin{center}
\begin{tabular}{l | l | l }
 & \ \ \ \ \ \ {\bf generic} &\ \ \ \ \  {\bf distinctive}\\
\midrule
 \textbf{dynamic}           & BabyBumps & CollegeBasketball \\
   & IAmA & Seahawks \\
       & Libertarian &  formula1 \\
       & australia & yugioh \\
\midrule
\textbf{consistent} & AdviceAnimals & Cooking \\

     & funny & Guitar \\
       & news & MakeupAddiction \\
       & pics & harrypotter \\
\end{tabular}
\vspace{-0.2cm}

\end{center}
\caption{Examples of communities on Reddit which occur at the extremes (top and bottom quartiles) of our typology. \label{tab:community_examples}
\vspace{-0.2cm}
}
\end{table}

\xhdr {Estimating linguistic measures} We estimate word frequencies 
$\prob_{c_t}  (w)$,
 and by extension each downstream measure, in a carefully controlled manner in order to ensure we capture robust and meaningful linguistic behaviour. %
First, we only consider top-level comments which are initial responses to a post, as the content of lower-level responses might reflect conventions of dialogue more than a community's high-level interests.
Next, in order to prevent a few highly active users from dominating our frequency estimates, we count each unique word {\em once} per user, ignoring successive uses of the same word
by the same user.
This ensures that our word-level characterizations are not  skewed by a small subset of 
highly active
contributors.\footnote{Understanding
 the role that highly active users \cite{hamilton_icwsm_2017} play %
in shaping a community's dynamics is an interesting direction for future work.}

In our subsequent analyses, we will only look at these measures computed over the {\em nouns} used in comments. 
In principle, our framework can be applied to any choice of vocabulary.
However, in the case of Reddit using nouns provides a convenient degree of interpretability.
We can easily understand the implication of a community preferentially mentioning a noun such as \textit{gamer} or \textit{feminist}, but interpreting the overuse of verbs or function words such as \textit{take} or \textit{of} is less straightforward. Additionally, in focusing on nouns we adopt the view emphasized in modern ``third wave'' accounts of sociolinguistic variation,
that stylistic variation is inseparable from 
topical content \cite{eckert2012three}.
In the case of online communities, the choice of what people choose to talk about serves as a primary signal 
of
 social identity. That said, a typology based on more purely stylistic differences is an interesting avenue for future work.

\begin{figure*}
\includegraphics[width=\textwidth]{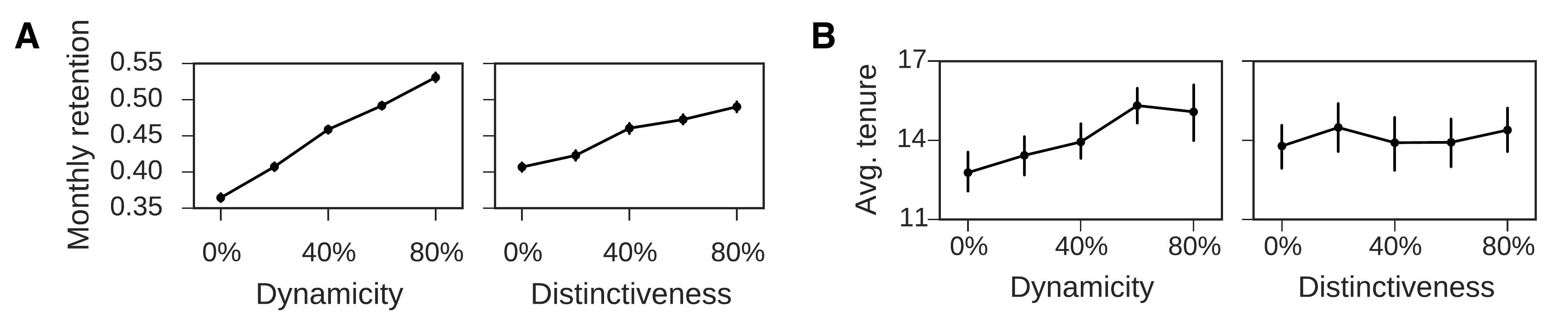}
\vspace{-0.75cm}
\caption{\textbf{A}: The monthly retention rate for communities differs drastically according to their position in our identity-based typology, with dynamicity being the strongest signal of higher user retention (x-axes bin community-months by percentiles; in all subsequent plots, error bars indicate 95\% bootstrapped confidence intervals). \textbf{B}: Dynamicity also correlates with long-term user retention, measured as the number of months the average user spends in the community; however, distinctiveness does not correlate with this longer-term variant of user retention.
\vspace{-0.37cm}
}
\label{fig:retention}
\end{figure*}

\xhdr{Accounting for rare words}
One complication when using measures such as PMI, which are based off of ratios of frequencies, is that estimates for very infrequent words could be overemphasized \cite{turney2003measuring}. Words that only appear a few times in a community tend to score at the extreme ends of our measures (e.g. as highly specific or highly generic), obfuscating the impact of more frequent words in the community. 
 To address this issue, we discard the long tail of infrequent words in our analyses, using only the top 5th percentile of words, by frequency within each $c_t$, to score comments and communities.\footnote{For the purposes of the present analyses, this method produces reasonable output that is robust to small variations in our choice of parameters. However, it would be fruitful in future work to consider other methods, e.g., \cite{monroe2008fightin}, for capturing linguistic variation.}

\xhdr{Typology output on Reddit}
The distribution of $\niche$ and $\dynam$ across Reddit communities is shown in Figure \ref{fig:figure0}.B, along with examples of communities at the extremes of our typology.
We find that 
interpretable groupings of communities emerge at various points within our axes.
For instance, highly distinctive and dynamic communities tend to focus on rapidly-updating interests like sports teams and games, while generic and consistent communities tend to be large ``link-sharing'' hubs where users generally post content with no clear dominating themes. More examples of communities at the extremes of our typology are shown in Table \ref{tab:community_examples}. 

We note that these groupings capture abstract properties of a community's content that go beyond its topic. For instance, our typology relates topically contrasting communities such as {\em yugioh} (which is about a popular trading card game) and {\em Seahawks} through the shared trait that their content is particularly distinctive. Additionally, the axes can clarify differences between topically similar communities: while {\em startrek} and {\em thewalkingdead} both focus on TV shows, {\em startrek} is less dynamic than the median community, while {\em thewalkingdead} is among the most dynamic communities, as the show 
was still airing during the years considered.

\section{Community identity and user retention}\label{sec:engagement}

\begin{figure*}
\centering
\includegraphics[scale=0.5]{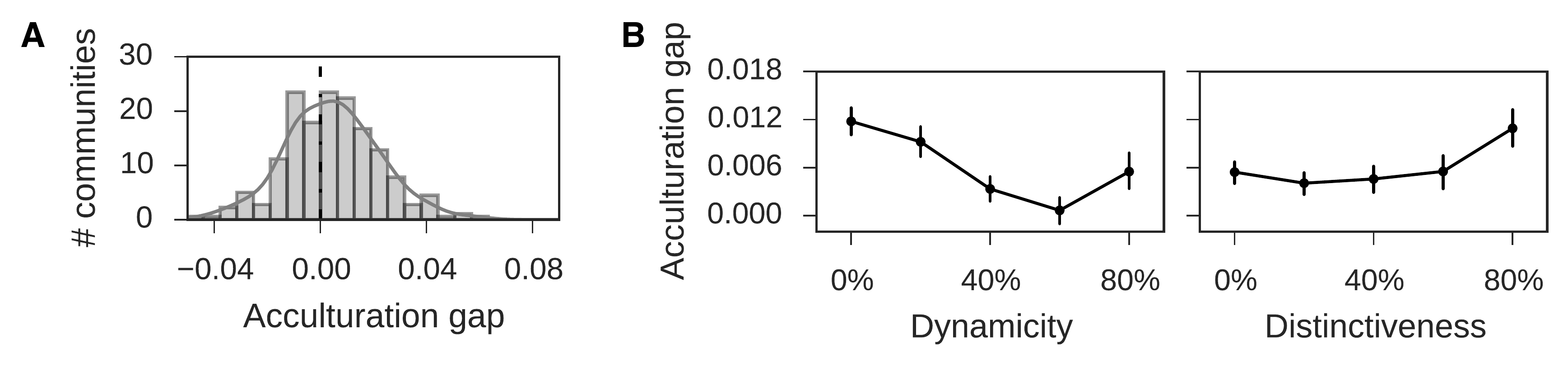}
\vspace{-0.2cm}
\caption{\textbf{A}: There is substantial variation in the direction and magnitude of the acculturation gap, which quantifies the extent to which established members of a community are linguistically differentiated from outsiders. Among 60\% of communities this gap is positive, indicating that established users match the community's language more than outsiders.
\textbf{B}: The size of the acculturation gap varies systematically according to how dynamic and distinctive a community is. Distinctive communities exhibit larger gaps; as do relatively stable, and very dynamic communities.\vspace{-0.2cm}
}
\label{fig:acculturation}
\end{figure*}

We have seen that our typology produces qualitatively satisfying groupings of communities according to the nature of their collective
 identity.
This section 
shows 
 that there is an informative and highly predictive relationship between a community's position in this typology and 
its user engagement patterns.
We find that communities with distinctive and dynamic identities have higher rates of user engagement, and further show that a community's position in our identity-based landscape holds important predictive information that is complementary to a strong activity baseline. 

In particular user retention is one of the most crucial aspects of engagement and is critical to community maintenance \cite{ren_encouraging_2012}.
We quantify how successful communities are at retaining users in terms of both short and long-term commitment. 
Our results indicate that 
 rates of user retention vary drastically, 
yet systematically according to how distinctive and dynamic a community is (Figure \ref{fig:figure0}).

We find a
strong, 
explanatory relationship between the temporal consistency of a community's identity and rates of user engagement:
dynamic communities that continually update and renew their discussion content tend to have far higher rates of user engagement. 
The relationship between distinctiveness and engagement is less universal, but still highly informative: niche communities tend to engender strong, focused interest from users at one particular point in time, though this does not necessarily translate into long-term retention. 

\subsection{Community-type and monthly retention}

We find that dynamic communities, such as \textit{Seahawks} or \textit{starcraft}, have substantially higher rates of monthly user retention than more stable communities  (Spearman's $\rho$ = 0.70, $p<$0.001, computed with community points averaged over months; Figure \ref{fig:retention}.A, left).  Similarly, more distinctive communities, like \textit{Cooking} and \textit{Naruto}, %
exhibit moderately higher monthly retention rates than more generic communities 
 (Spearman's $\rho$ = 0.33, $p<$0.001; Figure \ref{fig:retention}.A, right).

Monthly retention is formally defined as the proportion of users who contribute in month $t$ and then return to contribute again in month $t+1$.
Each monthly datapoint is treated as unique and the trends in Figure \ref{fig:retention} show 95\% bootstrapped confidence intervals, cluster-resampled at the level of subreddit \cite{field2007bootstrapping}, to account for differences in the number of months each subreddit contributes to the data.

Importantly, we find that in the task of predicting community-level user retention our identity-based typology 
holds additional predictive value  on top of strong baseline features based on community-size (\# contributing users) and activity levels (mean \# contributions per user), 
which are commonly used for churn prediction \cite{dror_churn_2012}. %
We compared out-of-sample predictive performance via leave-one-community-out cross validation using random forest regressors with ensembles of size 100, and otherwise default hyperparameters \cite{pedregosa_scikit_2011}.
A model predicting average monthly retention based on a community's average distinctiveness and dynamicity achieves an average mean squared error  ($\textrm{MSE}$)  of $0.0060$
and $R^2=0.37$,\footnote{We measure out-of-sample $R^2$ relative to a baseline that predicts the mean of the training data \cite{campbell2008predicting}.} 
while an analogous model predicting based on a community's size and average activity level (both log-transformed) achieves $\textrm{MSE}=0.0062$
and $R^2=0.35$.
The difference between the two models is not statistically significant
($p=0.99$, Wilcoxon signed-rank test).
However, combining features from both models results in a large and statistically significant improvement over each independent model ($\textrm{MSE}=0.0038$,  $R^2=0.60$, $p<0.001$ Bonferroni-corrected pairwise Wilcoxon tests).
These results indicate that our typology can explain variance in community-level retention rates, and provides 
   information beyond what is present in 
standard activity-based features. 

\subsection{Community-type and user tenure}
As with 
monthly
 retention, we find a strong positive relationship between a community's dynamicity and the  average number of months that a user will stay in that community (Spearman's $\rho$ = 0.41, $p<$0.001, computed over all community points; Figure \ref{fig:retention}.B, left).
This verifies that the short-term trend observed for monthly retention translates into longer-term engagement and suggests that long-term user retention 
might
 be strongly driven by the extent to which a community continually provides novel content. 
Interestingly, there is no significant relationship between distinctiveness and long-term engagement (Spearman's $\rho$ = 0.03, $p=$ 0.77; Figure \ref{fig:retention}.B, right).
Thus, while highly distinctive communities like \textit{RandomActsOfMakeup} may generate focused commitment from users over a short period of time, such communities are unlikely to retain long-term users unless they also have sufficiently dynamic content. 

To measure user tenures we focused on one slice of data (May, 2013)
 and measured how many months a user spends in each community, on average---the average number of months between a user's first and last comment in each community.\footnote{Analogous results hold for other reasonable choices of month.}
We have activity data up until May 2015, so the maximum tenure is 24 months in this set-up, which is exceptionally long relative to the average community member (throughout our entire data less than $3\%$ of users have tenures of more than 24 months in any community). 

\section{Community identity and acculturation}\label{sec:acculturation}

The previous section 
shows
 that there is a strong connection between the nature of a community's identity and 
its basic user engagement patterns.
In this section, we probe the relationship between a community's identity and how permeable, or accessible, it is to outsiders.

We measure this phenomenon using what we call the {\em acculturation gap}, which 
compares
 the extent to which engaged vs.\@ non-engaged users employ community-specific language. 
While previous work has found this gap to be large and predictive of future user engagement in two beer-review communities \cite{danescu-niculescu-mizil_no_2013}, we find that the size of the acculturation gap depends strongly on the nature of a community's identity, with the gap being most pronounced in stable, highly distinctive communities (Figure \ref{fig:acculturation}).

This finding has important implications for our understanding of online communities. 
Though many works have analyzed the dynamics of ``linguistic belonging'' in online communities \cite{cassell_language_2005,nguyen_language_2011,danescu-niculescu-mizil_no_2013,bamman2014gender}, our results suggest that the process of linguistically fitting in is highly contingent on the nature of a community's identity.
At one extreme, in generic communities like \textit{pics} or \textit{worldnews} there is no distinctive, linguistic identity for users to adopt.
To measure the acculturation gap for a community, we follow Danescu-Niculescu-Mizil et al \shortcite{danescu-niculescu-mizil_no_2013} and build ``snapshot language models'' (SLMs) for each community, which capture the linguistic state of a community at one point of time.\footnote{We use Katz-Backoff bigram language models with Good-Turing smoothing \cite{chen_empirical_1999} and vocabularies of size 50,000.}
Using these language models we can capture how linguistically close a particular utterance is to the community by measuring the cross-entropy of this utterance relative to the SLM: %
\newcommand{\slm}{\textrm{SLM}}
\begin{equation}
H(d, \slm_{c_t}) = \frac{1}{|d|}\sum_{b_i \in d}\slm_ {c_t}(b_i),
\end{equation}
where $\slm_{c_t}(b_i)$ is the probability assigned to bigram $b_i$ from comment $d$ in community-month $c_t$. 
We build the SLMs by randomly sampling 200 active users---defined as users with at least 5 comments in the respective community and month. For each of these 200 active users we select 5 random 10-word spans from 5 unique comments.\footnote{Using fixed-length spans controls for spurious length-effects \cite{danescu-niculescu-mizil_no_2013}; the same controls are used in the cross-entropy calculations.}
To ensure robustness and maximize data efficiency, we construct 100 SLMs for each community-month pair that has enough data, bootstrap-resampling from the set of active users. %

We compute a basic measure of the acculturation gap for a community-month $c_t$ as the relative difference of the cross-entropy of comments by users active in $c_t$ with that of singleton comments 
by {\em outsiders}---i.e., users who only ever commented once in $c$, but who are still active\footnote{Users must comment at least 5 times in a month to be considered active in Reddit.} in Reddit in general:
\begin{equation}\label{eq:acculturation}
A(c_t) = \frac{\mathbb{E}_{d \sim {\mathcal{V}_s}}[H(d,\slm_{c_t})]-\mathbb{E}_{d \sim {\mathcal{V}_a}}[H(d,\slm_{c_t})]}{\mathbb{E}_{d \sim \mathcal{V}_a}[H(d,\slm_{c_t})]}.
\end{equation}
$\mathcal{V}_s$ denotes the distribution over singleton comments, $\mathcal{V}_a$ denotes the distribution over comments from users active in $c_t$, and $\mathbb{E}$ the expected values of the cross-entropy over these respective distributions. %
For each bootstrap-sampled SLM we compute the cross-entropy of 50 comments by active users (10 comments from 5 randomly sampled active users, who were not used to construct the SLM) and 50 comments from randomly-sampled outsiders.

Figure \ref{fig:acculturation}.A shows that the acculturation gap varies substantially with how distinctive and dynamic a community is.
Highly distinctive communities have far higher acculturation gaps, while dynamicity exhibits a non-linear relationship: relatively stable communities have a higher linguistic `entry barrier', as do very dynamic ones. 
Thus, in communities like \textit{IAmA} (a general Q\&A forum) that are very generic, with content that is highly, but not extremely dynamic, outsiders are at no disadvantage
in matching the community's language.
In contrast, the acculturation gap is large in stable, distinctive communities like \textit{Cooking} that have consistent community-specific language.
The gap is also large in {\em extremely} dynamic communities like \textit{Seahawks},
which perhaps require more attention or interest on the part of active users to keep up-to-date with recent trends in content.

These results show that phenomena like the acculturation gap, %
which were previously observed in individual communities \cite{nguyen_language_2011,danescu-niculescu-mizil_no_2013}, cannot be easily generalized to a larger, heterogeneous set of communities. At the same time, we see that structuring the space of possible communities enables us to observe systematic patterns in how such phenomena vary. 
\section{Community identity and content affinity}\label{sec:divides}

Through the acculturation gap, we have shown that communities exhibit large yet systematic variations in their permeability to outsiders. We now turn to understanding the divide in commenting behaviour between outsiders and active community members at a finer granularity, by focusing on two particular ways in which such gaps might manifest among users: through different levels of engagement with {\em specific} content and with temporally {\em volatile} content. 

Echoing previous results, we find that community type mediates the extent and nature of 
the divide
 in content affinity. 
While in distinctive communities active members have a higher affinity for both community-specific content and for highly volatile content, the opposite is true for generic communities, where it is the outsiders who engage more with volatile content. %
We quantify these divides in content affinity by measuring differences in the language of the comments written by active users and outsiders.
Concretely,
for each community $c$, we define the {\em specificity gap} $\Delta \spec_c$ as the relative difference between the average specificity of comments authored by active members, and by outsiders, where these measures are macroaveraged over users. Large, positive $\Delta \spec_c$ then occur in communities where active users tend to engage with substantially more community-specific content than outsiders. 

We analogously define the {\em volatility gap} $\Delta \vol_c$ as the relative difference in volatilities of active member and outsider comments. Large, positive values of $\Delta \vol_c$ characterize communities where active users tend to have more volatile interests than outsiders, while negative values indicate communities where active users tend to have more stable interests.

We find that in 94\% of communities, $\Delta \spec_c > 0$, indicating (somewhat unsurprisingly) that in almost all communities, active users tend to engage with more community-specific content than outsiders. However, the magnitude of this divide can vary greatly: %
for instance, in \textit{Homebrewing}, which is dedicated to brewing beer, the divide is very pronounced ($\Delta \spec_c = $ 0.33) compared to \textit{funny}, a large hub where users share humorous content ($\Delta \spec_c =$ 0.011).

The nature of the volatility gap is comparatively more varied. In \textit{Homebrewing} ($\Delta \vol_c = $ 0.16), as in 68\% of communities, active users tend to write more volatile comments than outsiders ($\Delta \vol_c > $ 0). However, communities like \textit{funny} ($\Delta \vol_c = $ -0.16), where active users contribute relatively stable comments compared to outsiders ($\Delta \vol_c < $ 0), are also well-represented on Reddit. 

To understand whether these variations manifest systematically across communities, we examine the relationship between 
divides in content affinity and community type. In particular, following the intuition that %
active users have a relatively high affinity for a community's niche, we 
expect
 that the distinctiveness of a community will be a salient mediator of specificity and volatility gaps. Indeed, we find a strong correlation between a community's distinctiveness and its specificity gap (Spearman's $\rho =$ 0.34, $p <$ 0.001). 

We also find a strong correlation between distinctiveness and community volatility gaps (Spearman's $\rho =$ 0.53, $p <$ 0.001). In particular, we see that among the most {\em distinctive} communities (i.e., the top third of communities by distinctiveness),
active users tend to write more volatile comments than outsiders (mean $\Delta \vol_c =$ 0.098), while across the most {\em generic} communities (i.e., the bottom third), active users tend to write more {\em stable} comments (mean $\Delta \vol_c =$ -0.047, Mann-Whitney U test $p <$ 0.001).
The relative affinity of outsiders for volatile content in these communities  indicates that temporally ephemeral content might serve 
as an entry point into such a community, without necessarily engaging users in the long term.

\section{Further related work}

Our language-based typology and analysis of user engagement draws on and contributes to several distinct research threads, in addition to the many foundational studies cited in the previous sections. 

\xhdr{Multicommunity studies}
Our investigation of user engagement in multicommunity settings follows prior literature which has examined differences in user and community dynamics across various online groups, such as email listservs. %
Such studies have primarily related variations in user behaviour to structural features such as group size and volume of content
\cite{butler2001membership,jones2004information,backstrom2008preferential,kairam2012life}. In focusing on the linguistic content of communities, we extend this research by providing a {\em content-based} framework through which user engagement can be examined. 

Reddit has been a particularly useful setting for studying multiple communities in prior work. Such studies have mostly focused on characterizing how individual {\em users} engage across a multi-community platform \cite{tan_all_2015,hessel_science_2016}, or on specific user engagement patterns such as loyalty to particular communities \cite{hamilton_icwsm_2017}. 
We complement these studies by seeking to understand how features of {\em communities} can mediate a broad array of user engagement patterns within them.

\xhdr{Typologies of online communities}
Prior attempts to typologize online communities have primarily been qualitative and based on hand-designed categories, making them difficult to apply at scale.
These typologies often hinge on having some well-defined function the community serves, such as supporting a business or non-profit cause \cite{Porter:JournalOfComputerMediatedCommunication:2004}, which can be difficult or impossible to identify in massive, anonymous multi-community settings.
Other typologies emphasize differences in communication platforms and other functional requirements \cite{preece2001sociability,StanoevskaSlabeva:SystemSciences:2001}, which are important but preclude analyzing differences between communities within the same multi-community platform. Similarly, previous computational methods of characterizing multiple communities have relied on the presence of markers such as affixes in community names \cite{hessel_science_2016}, or platform-specific affordances such as evaluation mechanisms \cite{lee2016beyond}.

Our typology is also distinguished from community detection techniques that rely on structural or functional categorizations \cite{Leskovec:2008:SPC:1367497.1367591,Yang:KnowledgeAndInformationSystems:2013}.
While the focus of those studies is to identify and characterize sub-communities within a larger social network, our typology provides a characterization of pre-defined communities based on the nature of their 
 identity.

\xhdr{Broader work on collective identity}
Our focus on community identity dovetails with a long line of research on collective identity and user engagement, in both online and offline communities \cite{allen_affective_1996,tajfel_social_2010,ren_encouraging_2012}.
These studies focus on individual-level psychological manifestations of collective (or social) identity, and their relationship to user engagement \cite{allen_affective_1996,meyer_affective_2002,utz_social_2003,ren_applying_2007}.

In contrast, we seek to characterize community identities at an aggregate level and in an interpretable manner, with the goal of systematically organizing the diverse space of online communities.
Typologies of this kind are critical to these broader, social-psychological studies of collective identity: they allow researchers to systematically analyze how the psychological manifestations and implications of collective identity vary across diverse sets of communities. 

\section{Conclusion and future work}
Our current understanding of engagement patterns in online communities is patched up from glimpses offered by several disparate studies focusing on a few individual communities.  This work calls into attention the need for a method to systematically reason about similarities and differences across communities.  
By proposing a way to structure the multi-community space, we find not only that radically contrasting engagement patterns emerge in different parts of this space, but also that this variation can be at least partly explained by the type of identity each community fosters.

Our choice in this work is to structure the multi-community space according to a typology based on community identity, as reflected in language use.  We show that this effectively explains cross-community variation of three different user engagement measures---retention, acculturation and content affinity---and complements measures based on activity and size with additional interpretable information.   For example, we find that in niche communities established members are more likely to engage with volatile content than outsiders, while the opposite is true in generic communities.  Such insights can be useful for community maintainers seeking to understand engagement patterns in their own communities.

One main area of future research is to examine the {\em temporal dynamics}  in the multi-community landscape. By averaging our measures of distinctiveness and dynamicity across time, our present study treated 
 community identity as a static property. However, as communities experience internal changes and respond to external events, we can expect the nature of their identity to shift as well. For instance, the relative consistency of {\em harrypotter} may be disrupted by the release of a new novel, while {\em Seahawks} may foster different identities during and between football seasons. Conversely, a community's type may also mediate the impact of new events. Moving beyond a static view of community identity could enable us to better understand how temporal phenomena such as linguistic change manifest across different communities, and also provide a more nuanced view of user engagement---for instance, are communities more welcoming to newcomers at certain points in their lifecycle?  

Another important avenue of future work is to explore other ways of mapping the landscape of online communities.  For example, combining structural properties of communities \cite{Leskovec:2008:SPC:1367497.1367591} with topical information \cite{hessel_science_2016} and with our identity-based measures could further characterize and explain variations in user engagement patterns.  Furthermore, extending the present analyses to even more diverse communities supported by different platforms (e.g.,  GitHub, StackExchange, Wikipedia) could enable the characterization of more complex behavioral patterns such as collaboration and altruism, which become salient in different multicommunity landscapes.

\section*{Acknowledgements} 
The authors thank Liye Fu, Jack Hessel, David Jurgens and Lillian Lee for their helpful comments.
This research has been supported in part by a Discovery and Innovation Research Seed Award from the Office of the Vice Provost for Research at Cornell, NSF
CNS-1010921,     
IIS-1149837, IIS-1514268
NIH BD2K,
ARO MURI, DARPA XDATA,
DARPA SIMPLEX, DARPA NGS2,
Stanford Data Science Initiative,
SAP Stanford Graduate Fellowship, NSERC PGS-D,
Boeing,          
Lightspeed,			       
and Volkswagen.
\bibliographystyle{aaai}
\bibliography{taxonomy}

\end{document}